\documentclass[twocolumn,superscriptaddress,amsmath,amssymb,aps,prb]{revtex4-2}
\usepackage{graphicx}
\usepackage{CJKutf8}
\usepackage{dcolumn}
\usepackage{amsmath,bm}
\usepackage{epstopdf}
\usepackage[mathlines]{lineno}
\usepackage{color}
\usepackage[colorlinks=true,linkcolor=blue,citecolor=magenta,urlcolor=cyan]{hyperref}
\usepackage{orcidlink}

\begin{document}
\title{Hole subband dispersions and strong ‘spin’-orbit coupling in a cylindrical Ge nanowire}
\author{Rui\! Li~(\begin{CJK}{UTF8}{gbsn}李睿\end{CJK})\,\orcidlink{0000-0003-1900-5998}}
\email{ruili@ysu.edu.cn}
\affiliation{Hebei Key Laboratory of Microstructural Material Physics, School of Science, Yanshan University, Qinhuangdao 066004, China}

\begin{abstract}
{\color{red}}
Quasi-one-dimensional hole gas is achievable in a semiconductor Ge nanowire. The lowest two subband dispersions of the hole gas are just two shifted parabolic curves with an anticrossing at $k_{z}=0$. This peculiar low-energy subband structure manifests the existence of a strong `spin' (pseudo spin)-orbit coupling. Based on the Luttinger-Kohn Hamiltonian in the axial approximation, we show two sets of combined dispersions that not only isolated from each other but also with strong `spin'-orbit coupling are obtainable in the presence of strong magnetic field. Realistic calculations are performed for three representative nanowire growth directions [001], [111], and [110]. These results are further confirmed via constructing the low-energy effective Hamiltonian of the hole gas.  We also calculate the external electric field induced spin splitting for comparison with the magnetic field induced spin splitting.

\end{abstract}
\date{\today}
\maketitle

\section{Introduction}
{\color{red}}
The valence bands of bulk semiconductors are composed of a heavy-hole, a light-hole, and a spin-orbit split-off bands~\cite{bir1974symmetry,winkler2003spin,sun2009strain}. The heavy-hole and light-hole bands touch with each other at the valence band top, i.e., the $\Gamma$  point of the Brillouin zone. The spin-orbit split-off band is separated from the heavy-hole and light-hole bands by a spin-orbit energy gap~\cite{KANE1957249}. The effective Hamiltonian describing the valence bands near the $\Gamma$ point was developed by Luttinger and Kohn~\cite{PhysRev.97.869}. When the spin degree of freedom is included, the Luttinger-Kohn Hamiltonian is a $6\times6$ matrix, which is complicated by itself even before its applications to the low-dimensional nanostructures~\cite{BANGERT1985363,PhysRevB.48.4964,PhysRevB.46.4110,PhysRevB.53.10858}. Fortunately, when the spin-orbit energy gap is much larger than the other energy scales in the nanostructures, we can safely use the $4\times4$ Luttinger-Kohn Hamiltonian instead~\cite{PhysRevB.31.888,RASHBA1988175,PhysRevB.36.5887,PhysRevB.43.9649}. The $4\times4$ Luttinger-Kohn Hamiltonian describes the heavy-hole and light-hole bands only, and is regarded as a minimal bulk hole model~\cite{PhysRev.97.869,PhysRev.102.1030}.

Holes in traditional quasi-two-dimensional semiconductor structures, such as metal-oxide-semiconductor structure~\cite{PhysRevB.58.9941}, heterojunction~\cite{ando1985hole,PhysRevB.32.3712,1073186,PhysRevB.62.4245,PhysRevB.65.155303,PhysRevB.95.075305}, and quantum well~\cite{ando1985hole,PhysRevB.32.5138,PhysRevB.36.5887,RASHBA1988175}, were well studied (see also the reviews~\cite{Schaffler:1997vc,WU201061,Scappucci:2021vk}). While holes in quasi-one-dimensional nanostructures had not attracted much attention, until recently their potential applications in field effect transistor~\cite{Xiang:2006uq} and spin quantum computation~\cite{Pribiag:2013vb} were recognized. Quasi-one-dimensional hole gas can be achieved in various nanowire structures, such as Ge/Si core/shell nanowire~\cite{Lu10046}, Ge hut wire~\cite{Gao2020AM}, and InSb nanowire~\cite{Pribiag:2013vb} etc. Note that current experiments mainly used undoped nanowire, the two ends of which are connected to metal electrodes~\cite{10.1063/1.4963715}. The nanowire can be populated with holes by a back gate, i.e., a ${\rm p}^{++}$ doped Si layer lying below the nanowire~\cite{10.1063/1.4963715,PhysRevB.94.041411}. 

Holes in the nanowire move freely only in the longitudinal direction. In order to understand the various hole properties in quasi-one-dimension, the first step is to understand the hole subband dispersions caused by the size quantization of the nanowire. Using the Luttinger-Kohn Hamiltonian in the spherical approximation~\cite{PhysRev.102.1030}, Sercel and Vahala obtained the transcendental equation determining the subband dispersions as early as 1990~\cite{PhysRevB.42.3690}. Sweeny et al. even gave an equivalent transcendental equation two years earlier using a coordinate free method~\cite{sweeny1988hole}. In 2024, Li generalized the transcendental equations to the case of the axial approximation~\cite{Li-lirui:2024we}. The axial approximation is superior to the spherical approximation for it can describe the band anisotropy to some extent~\cite{PhysRevB.9.4184,PhysRevB.20.686}. In particular, the axial approximation gives rise to highly accurate subband dispersions for both growth directions [001] and [111]~\cite{winkler2003spin}.

The lowest two subband dispersions obtained are just two shifted parabolic curves with an anticrossing at $k_{z}=0$~\cite{PhysRevB.84.195314,RL2021,RL2022a,RL2023b,RL2023c}. This result indicates the hole gas is naturally strong `spin' (pseudo spin)-orbital coupled, where the `spin' is introduced to describe the two shifted parabolic curve~\cite{PhysRevB.84.195314,RL2023b}. Note that each dispersion line is spin degenerate. Using a strong magnetic field to lift this spin degeneracy, we obtain two sets of combined dispersions, and each set is strong `spin'-orbital coupled~\cite{RL2023b}. The magnetic field can be applied either longitudinally or transversely. A strong transverse external electric field can also lift the spin degeneracy at $k_{z}\neq0$ due to the Rashba effect~\cite{bychkov1984oscillatory,PhysRevB.62.4245,PhysRevB.84.195314,PhysRevB.97.235422,PhysRevB.105.075308,PhysRevB.106.235408,PhysRevLett.119.126401}, strong hole spin-orbit coupling, instead of `spin'-orbit coupling, is achievable in this case. 

The strong `spin' or spin-orbit coupling achieved either by strong magnetic field or by strong electric field has many potential applications. The strong `spin' or spin-orbit coupling can be used to facilitate the hole spin manipulation in an oscillating electric field~\cite{Wang:2022tm,PhysRevB.99.115317,PhysRevB.103.045305,PhysRevB.104.235304,RL2023a}, to achieve the strong spin-photon interaction in a quantum-dot-cavity system~\cite{PhysRevB.88.241405,RL_2024a,Li:2018ab}, to achieve the topological superconductivity in hybrid semiconductor-superconductor structures~\cite{PhysRevB.90.195421,PhysRevB.108.155433,PhysRevB.109.035433}. Spin relaxation induced by lattice phonons~\cite{Hu:2012ws,PhysRevB.87.161305} and spin dephasing induced by charge noise~\cite{RL2018c,RL2020,RL_2024a} can also be mediated by this spin-orbit coupling.

This paper is organized as follows. In Sec.~\ref{sec_LK}, the Luttinger-Kohn Hamiltonian and two types of approximation are introduced. In Sec.~\ref{sec_effmassH}, the symmetry of the effective mass Hamiltonian is analyzed. In Sec.~\ref{sec_traneq}, the transcendental equations determining the energy spectrum are given. In Sec.~\ref{sec_subbands}, subband dispersions for growth direction [110] are calculated. In Sec.~\ref{sec_splittingmagnetic}, spin splittings induced by external magnetic field are calculated. In Sec.~\ref{sec_effhamiltonian}, a series of lower-energy effective Hamiltonians are constructed. We finally give a summary in Sec.~\ref{sec_summary}.

\section{\label{sec_LK}Luttinger-Kohn Hamiltonian}
{\color{red}}
In this paper, we adopt the $4\times4$ Luttinger-Kohn Hamiltonian~\cite{PhysRev.97.869,PhysRev.102.1030} to describe the hole kinetic energy in the Ge nanowire. The $4\times4$ Luttinger-Kohn Hamiltonian is accurate enough because the typical quantization energy (about tens of meV~\cite{PhysRevB.84.195314}) in the nanowire is much less than the spin-orbit energy gap (about 297 meV~\cite{madelung2004semiconductors}). It is reasonable to neglect the effects of the spin-orbit split-off band. When the wave vectors are along the cubic axes of the crystal, i.e., $k_{x}\parallel[100]$, $k_{y}\parallel[010]$, and $k_{z}\parallel[001]$, the Luttinger-Kohn Hamiltonian reads~\cite{PhysRev.97.869,PhysRev.102.1030}
\begin{eqnarray}
H_{\rm LK}&=&\frac{\hbar^{2}}{2m_{0}}\Big[\Big(\gamma_{1}+\frac{5\gamma_{2}}{2}\Big){\boldsymbol k}^{2}-2\gamma_{2}\left(k^{2}_{x}J^{2}_{x}+k^{2}_{y}J^{2}_{y}+k^{2}_{z}J^{2}_{z}\right)\nonumber\\
&&-4\gamma_{3}\left(\{k_{x},k_{y}\}\{J_{x},J_{y}\}+{\rm c.p.}\right)\Big],\label{eq_LKoriginal}
\end{eqnarray}
where $m_{0}$ is the free electron mass, $\gamma_{1}=13.35$, $\gamma_{2}=4.25$ and $\gamma_{3}=5.69$ are Luttinger parameters of semiconductor Ge~\cite{PhysRevB.4.3460}, $J_{x,y,z}$ are standard spin-3/2 matrices (for details see appendix~\ref{Appendix_A}), {\rm c.p.} denotes cyclic permutations, and $\{A,B\}=(AB+BA)/2$. 

If the wave vectors $k_{x,y,z}$ are not along the crystal cubic axes, we need to obtain the corresponding Luttinger-Kohn Hamiltonian using coordinate transformation~\cite{PhysRevB.43.9856,PhysRevB.97.235422}. The calculations of the subband dispersions in the nanowire are indeed complicated, some studies even use purely numerical methods~\cite{doi:10.1063/1.4929412,doi:10.1063/1.4972987,SAIDI2020136872,PhysRevB.105.245303,PhysRevB.73.165319}, the primitive Luttinger-Kohn Hamiltonian (\ref{eq_LKoriginal}) is rarely used in analytical studies. As such, two types of approximation are usually applied to Hamiltonian (\ref{eq_LKoriginal}).

The simplest and the most frequently used approximation is the spherical approximation~\cite{PhysRevB.8.2697,PhysRevB.40.8500}. Simply replacing $\gamma_{2,3}$ with $\gamma_{s}$ in Eq.~(\ref{eq_LKoriginal}), one obtains the Luttinger-Kohn Hamiltonian in the spherical approximation
\begin{equation}
H^{\rm sp}_{\rm LK}=\frac{\hbar^{2}}{2m_{0}}\Big[\Big(\gamma_{1}+\frac{5}{2}\gamma_{s}\Big){\boldsymbol k}^{2}-2\gamma_{s}({\boldsymbol k}\cdot{\boldsymbol J})^{2}\Big].\label{eq_LKsph}
\end{equation}
In most applications, the new Luttinger parameter $\gamma_{s}$ is set to $\gamma_{s}=(2\gamma_{2}+3\gamma_{3})/5$~\cite{PhysRevB.8.2697,PhysRevB.84.195314}. Note that in Hamiltonian (\ref{eq_LKsph}), the wave vectors $k_{x,y,z}$ are not necessary along special crystal directions, as long as they are mutually perpendicular to each other. Hence, the spherical approximation is unable to reflect the dispersion anisotropy.

\begin{figure}
\includegraphics[width=5cm]{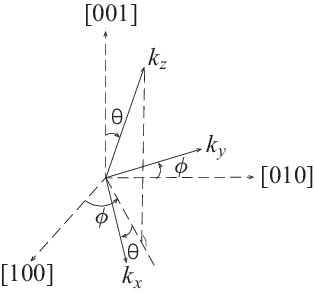}
\caption{\label{fig_coordinate}The orientations of the coordinate axes $k_{x,y,z}$ relative to the crystal cubic axes are described by two angles $\theta$ and $\phi$. Reproduced from~\cite{Li-lirui:2024we}.}
\end{figure}

The other approximation is called the axial approximation~\cite{PhysRevB.9.4184,PhysRevB.20.686}, which is regarded as a more advanced approximation in comparison with the spherical one. The axial approximation can describe the dispersion anisotropy to some extent through its directionally dependent Luttinger parameters. The Luttinger-Kohn Hamiltonian in the axial approximation reads~\cite{winkler2003spin}
\begin{eqnarray}
H^{\rm ax}_{\rm LK}&=&\frac{\hbar^{2}}{2m_{0}}\Big[\gamma_{1}{\boldsymbol k}^{2}+\tilde{\gamma}_{1}(k^{2}_{x}+k^{2}_{y}-2k^{2}_{z})\Big(J^{2}_{z}-\frac{5}{4}\Big)\nonumber\\
&&-2\tilde{\gamma}_{2}\big(\{k_{z},k_{+}\}\{J_{z},J_{-}\}+\{k_{z},k_{-}\}\{J_{z},J_{+}\}\big)\nonumber\\
&&-\frac{\tilde{\gamma}_{3}}{2}(k^{2}_{+}J^{2}_{-}+k^{2}_{-}J^{2}_{+})\Big],\label{eq_LKaxial}
\end{eqnarray}
where $k_{\pm}=k_{x}\pm\,ik_{y}$, $J_{\pm}=J_{x}\pm\,iJ_{y}$, and $\tilde{\gamma}_{1,2,3}$ are Luttinger parameters depending on the orientations of the coordinate axes $k_{x,y,z}$ in the crystal. The relationship between $\tilde{\gamma}$ and $\gamma$ reads~\cite{winkler2003spin}
\begin{eqnarray}
\tilde{\gamma}_{1}&=&(1-\zeta)\gamma_{2}+\zeta\gamma_{3},\nonumber\\
\tilde{\gamma}_{2}&=&2\zeta\gamma_{2}/3+(1-2\zeta/3)\gamma_{3},\nonumber\\
\tilde{\gamma}_{3}&=&(3-\zeta)\gamma_{2}/6+(3+\zeta)\gamma_{3}/6,\label{eq_gamma}
\end{eqnarray}
where $\zeta=\sin^{2}\theta[3-(3/8)(7+\cos4\phi)\sin^{2}\theta]$, with $\theta$ being the azimuthal angle and $\phi$ being the polar angle of the $k_{z}$ axis with respect to the [001] direction (see Fig.~\ref{fig_coordinate})~\cite{winkler2003spin}.

Hamiltonian $H^{\rm sp}_{\rm LK}$ commutes with both ${\boldsymbol F}^{2}$ and $F_{z}$~\cite{PhysRevB.42.3690}, where ${\boldsymbol F}={\boldsymbol L}+{\boldsymbol J}$ is the total angular momentum, with ${\boldsymbol L}$ being the orbital angular momentum. Hamiltonian $H^{\rm ax}_{\rm LK}$ commutes with $F_{z}$ only~\cite{winkler2003spin}. The conservation of $F_{z}$ leads to exact solution of the subband dispersions in the nanowire when either $H^{\rm sp}_{\rm LK}$ or $H^{\rm ax}_{\rm LK}$ is used~\cite{PhysRevB.42.3690,sweeny1988hole,Li-lirui:2024we}. Also, the axial approximation is most accurate for high-symmetry nanowire growth directions [001] ($\theta=\phi=0$)~\cite{winkler2003spin,PhysRevB.52.11132} and [111] ($\theta=\arccos(1/\sqrt{3})$, $\phi=\pi/4$)~\cite{winkler2003spin,PhysRevB.85.235308,PhysRevB.108.165301}.

\section{\label{sec_effmassH}Effective mass Hamiltonian and symmetry analysis}
{\color{red}}
\begin{figure}
\includegraphics{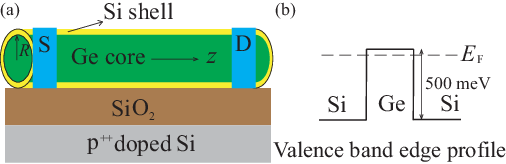}
\caption{\label{fig_GeSiNanowire}(a) The setup used to achieve quasi-one-dimensional hole gas. The ${\rm p}^{++}$ doped Si substrate is covered with ${\rm SiO}_{2}$, on which a Ge/Si core/shell nanowire is deposited. The contacts between the nanowire and the source/drain electrodes are ohmic. (b) The valence band edge profile of the Ge/Si core/shell heterostructure. The dashed line indicates the position of the Fermi level.}
\end{figure}

Quasi-one-dimensional hole gas is experimentally achievable in Ge/Si core/shell nanowire heterostructure~\cite{Lu10046,PhysRevLett.101.186802} and Ge hut wire~\cite{Watzinger:2016aa,Gao2020AM,Zhang:2021wj}. A typical experimental setup is shown in Fig.~\ref{fig_GeSiNanowire}(a). The undoped Ge core can be populated with holes from the ohmic contacts when negative voltage is applied to the ${\rm p}^{++}$ doped Si substrate~\cite{10.1063/1.4963715,PhysRevB.93.121408}. The Si shell serves as a strong confinement to the hole gas due to the large valence band offset between Ge and Si (about 500 meV~\cite{PhysRevB.34.5621}, see also Fig.~\ref{fig_GeSiNanowire}(b)). Since the quantization energy in the nanowire is much less than this band offset, the transverse confinement can be treated as a hard wall. 

We consider a cylindrical Ge nanowire, the axis of which is defined as the $z$ axis (see Fig.~\ref{fig_GeSiNanowire}(a)). Certainly, the nanowire axis, i.e., the nanowire growth direction, can point along various crystal directions (see Fig.~\ref{fig_coordinate}). The effective mass Hamiltonian of the hole gas reads
\begin{equation}
H_{0}=H^{\rm ax}_{\rm LK}({\boldsymbol k}\rightarrow-i\nabla)+V(r),\label{eq_efmassHamil}
\end{equation}
where
\begin{equation}
V(r)=\left\{\begin{array}{cc}0,~&~r<R,\\
\infty,~&~r>R,\end{array}\right.
\end{equation}
with $R$ being the radius of the Ge core. Here we have chosen to study in a cylindrical coordinate system, where $r=\sqrt{x^{2}+y^{2}}$, $\varphi=\arctan(y/x)$, and $z=z$. Also, this paper focuses on the analytical solution to the effective mass model, such that the axial Luttinger-Kohn Hamiltonian is employed. Now, let us analyze the symmetries of the model~(\ref{eq_efmassHamil}) and their important consequence: the spin degeneracy in the subband dispersions~\cite{RL2021}. 

First, both $k_{z}$ and $F_{z}=-i\partial_{\varphi}+J_{z}$ commute with Hamiltonian (\ref{eq_efmassHamil}), such that them can be used to classify the eigenfunctions. For a general value of $F_{z}=m+1/2$ ($m=0,\pm1,\pm2\ldots$), the eigenfunction can be written as~\cite{RL2021,Li-lirui:2024we} 
\begin{equation}
\Psi_{k_{z},F_{z}}(r,\varphi,z)=\left(\begin{array}{c}\Psi_{1}(r)e^{i(m-1)\varphi}\\\Psi_{2}(r)e^{im\varphi}\\\Psi_{3}(r)e^{i(m+1)\varphi}\\\Psi_{4}(r)e^{i(m+2)\varphi}\end{array}\right)e^{ik_{z}z},\label{Eq_eigenfunc}
\end{equation}
where $\Psi_{1,2,3,4}(r)$ are four components of the eigenfunction. Note that they are functions of the coordinate $r$ only.

Second, Hamiltonian (\ref{eq_efmassHamil}) is time-reversal invariant. The time-reversal operator for a spin-3/2 system reads
\begin{equation}
T=\Gamma_{1}\Gamma_{3}K,
\end{equation}
where $K$ is the usual complex conjugate operator, and operators $\Gamma_{1,3}$ are given in appendix {\ref{Appendix_A}}~\cite{PhysRevLett.91.186402}. One can verify the following relations $T^{2}=-1$, $T{\bf J}T^{-1}=-{\bf J}$, and $T{\bf k}T^{-1}=-{\bf k}$, such that the time-reversal symmetry $TH_{0}T^{-1}=H_{0}$ becomes evident. Acting the time-reversal operator $T$ on the eigenfunction (\ref{Eq_eigenfunc}), we have
\begin{equation}
\Psi_{-k_{z},-F_{z}}(r,\varphi,z)=\left(\begin{array}{c}-\Psi^{*}_{4}(r)e^{-i(m+2)\varphi}\\\Psi^{*}_{3}(r)e^{-i(m+1)\varphi}\\-\Psi^{*}_{2}(r)e^{-im\varphi}\\\Psi^{*}_{1}(r)e^{-i(m-1)\varphi}\end{array}\right)e^{-ik_{z}z}.\label{Eq_eigenfunT}
\end{equation}
At this stage, the Kramer's degeneracy only gives rise to $E(k_{z},F_{z})=E(-k_{z},-F_{z})$~\cite{kittel1963quantum}.

Third, Hamiltonian (\ref{eq_efmassHamil}) is space-inversion invariant. The space-inversion operator $I$ sets the coordinate ${\bf r}=(r,\varphi,z)$ to $-{\bf r}=(r,\varphi+\pi,-z)$, e.g., $I\Psi({\bf r})=\Psi(-{\bf r})$. One can verify $IH_{0}I^{-1}=H_{0}$. Acting the space-reversal operator $I$ on the eigenfunction (\ref{Eq_eigenfunT}), we have
\begin{equation}
\Psi_{k_{z},-F_{z}}(r,\varphi,z)=\left(\begin{array}{c}\Psi^{*}_{4}(r)e^{-i(m+2)\varphi}\\\Psi^{*}_{3}(r)e^{-i(m+1)\varphi}\\\Psi^{*}_{2}(r)e^{-im\varphi}\\\Psi^{*}_{1}(r)e^{-i(m-1)\varphi}\end{array}\right)e^{ik_{z}z}.\label{Eq_eigenfunc2}
\end{equation}

Since both $\Psi_{k_{z},F_{z}}(r,\varphi,z)$ and $\Psi_{k_{z},-F_{z}}(r,\varphi,z)$ correspond to the same energy eigenvalue, while they must represent two distinct quantum states, i.e., the $F_{z}$ values are different in these two states, we thus have the spin degeneracy in the hole subband dispersions 
\begin{equation}
E(k_{z},F_{z})=E(k_{z},-F_{z}).
\end{equation} 
Here, the two quantized directions of the hole spin are indicated by the $'\pm'$ signs before $F_{z}$. Hence, the spin degeneracy is a direct consequence of the combined effect of the time-reversal symmetry and space-inversion symmetry~\cite{kittel1963quantum,winkler2003spin}.

\section{\label{sec_traneq}The transcendental equations}
{\color{red}}
The potential $V(r)$ in Hamiltonian (\ref{eq_efmassHamil}) is a hard-wall potential, it is zero inside the wall and infinite outside the wall. In this sense, the method of solving model (\ref{eq_efmassHamil}) should be similar to that of solving an infinite square-well in a quantum mechanics textbook~\cite{landau1965quantum}, or more precisely to that of solving a  hard-wall quantum dot with spin-orbit coupling~\cite{Bulgakov2001,PhysRevB.70.115316}. The solving procedure can be summarized as follows. We first obtain both the bulk spectrum and the corresponding bulk wavefunctions governed by $H^{\rm ax}_{\rm LK}$. We then write the eigenfunction of Hamiltonian (\ref{eq_efmassHamil}) as a linear combination of the bulk wavefunctions. We finally let the eigenfunction vanish at the boundary $r=R$, such that a series of transcendental equations determining the subband energies are obtainable.

Both the bulk spectrum and the corresponding bulk wavefunctions are explicitly derived in Ref.~\cite{Li-lirui:2024we}. In writing the eigenfunction as a linear combination of the bulk wavefunctions, we should carefully take into account of the region division of the energy $E$. The selection of the bulk wavefunctions varies with the energy region, especially for growth direction [001]. The hard-wall boundary condition of the eigenfunction gives rise to the following transcendental equations~\cite{Li-lirui:2024we} 
\begin{widetext}
\begin{equation}
\left|\begin{array}{cccc}
\frac{2i\tilde{\gamma}_{2}k_{z}}{\tilde{\gamma}_{3}\mu_{1}}J_{m-1}(\mu_{1}R)&\frac{\tilde{\gamma}_{1}(\mu^{2}_{1}-2k^{2}_{z})\pm\chi_{\mu_{1}}}{\sqrt{3}\tilde{\gamma}_{3}\mu^{2}_{1}}J_{m-1}(\mu_{1}R)&\frac{2i\tilde{\gamma}_{2}k_{z}}{\tilde{\gamma}_{3}\mu_{2}}J_{m-1}(\mu_{2}R)&\frac{\tilde{\gamma}_{1}(\mu^{2}_{2}-2k^{2}_{z})-\chi_{\mu_{2}}}{\sqrt{3}\tilde{\gamma}_{3}\mu^{2}_{2}}J_{m-1}(\mu_{2}R)\\
\frac{-\tilde{\gamma}_{1}(\mu^{2}_{1}-2k^{2}_{z})\pm\chi_{\mu_{1}}}{\sqrt{3}\tilde{\gamma}_{3}\mu^{2}_{1}}J_{m}(\mu_{1}R)&-\frac{2i\tilde{\gamma}_{2}k_{z}}{\tilde{\gamma}_{3}\mu_{1}}J_{m}(\mu_{1}R)&\frac{-\tilde{\gamma}_{1}(\mu^{2}_{2}-2k^{2}_{z})-\chi_{\mu_{2}}}{\sqrt{3}\tilde{\gamma}_{3}\mu^{2}_{2}}J_{m}(\mu_{2}R)&-\frac{2i\tilde{\gamma}_{2}k_{z}}{\tilde{\gamma}_{3}\mu_{2}}J_{m}(\mu_{2}R)\\
0&J_{m+1}(\mu_{1}R)&0&J_{m+1}(\mu_{2}R)\\
J_{m+2}(\mu_{1}R)&0&J_{m+2}(\mu_{2}R)&0
\end{array}\right|=0,\label{eq_transc}
\end{equation}
\end{widetext}
where $J_{m}(\mu\,r)$ is the $m$-order Bessel function~\cite{zhuxi_wang} and
\begin{equation}
\mu_{1,2}=\left(\frac{-b\mp\sqrt{b^{2}-4ac}}{2a}\right)^{1/2},\label{eq_mu}
\end{equation}
with
\begin{eqnarray}
a&=&\gamma^{2}_{1}-\tilde{\gamma}^{2}_{1}-3\tilde{\gamma}^{2}_{3},\nonumber\\
b&=&2\big[\gamma^{2}_{1}k^{2}_{z}+2(\tilde{\gamma}^{2}_{1}-3\tilde{\gamma}^{2}_{2})k^{2}_{z}-2\gamma_{1}m_{0}E/\hbar^{2}\big],\nonumber\\
c&=&\gamma^{2}_{1}k^{4}_{z}-4\tilde{\gamma}^{2}_{1}k^{4}_{z}-4\gamma_{1}k^{2}_{z}m_{0}E/\hbar^{2}+4m^{2}_{0}E^{2}/\hbar^{4}.
\end{eqnarray}
The `$\pm$' signs in Eq.~(\ref{eq_transc}) correspond to two energy regions~\cite{Li-lirui:2024we}. The energy eigenvalue $E$ is the only unknown in Eq.~(\ref{eq_transc}). We can obtain the hole subband dispersions $E(k_{z})$ via numerically solving Eq.~(\ref{eq_transc}). Once an eigenvalue $E$ is obtained, the corresponding eigenfunction can also be obtained (see appendix~\ref{Appendix_b})~\cite{RL2021}. Note that in most cases, the plus sign `+' transcendental equation (\ref{eq_transc}) alone already gives the subband energies in the most interested range of the wave vector $k_{z}$.

Both the energy eigenvalues and eigenfunctions at the wave vector site $k_{z}=0$ are important in constructing the low-energy effective Hamiltonian of the hole gas (see Sec.~\ref{sec_effhamiltonian}). Several studies have also studied the hole $g$-factor at this site by using the spherical approximation~\cite{PhysRevB.76.073313,PhysRevB.78.033307,PhysRevB.79.155323}. When $k_{z}=0$, the plus sign `+' transcendental equation (\ref{eq_transc}) can be reduced to two independent equations~\cite{Li-lirui:2024we}. The eigenfunctions now always have two vanishing components, i.e., either $\Psi_{1,3}(r)$ or $\Psi_{2,4}(r)$ are zero (see appendix~\ref{Appendix_b})~\cite{PhysRevB.79.155323,PhysRevB.84.195314,RL2021}.

If we want to use the spherical approximation instead of the axial approximation,  we just need to replace $\tilde{\gamma}_{1,2,3}$ with $\gamma_{s}$ in Eq.~(\ref{eq_transc}). The result of Sercel and Vahala~\cite{PhysRevB.42.3690,PhysRevB.44.5681} is nicely recovered in this case.

\section{\label{sec_subbands}Subband dispersions}
{\color{red}}
\begin{figure}
\includegraphics{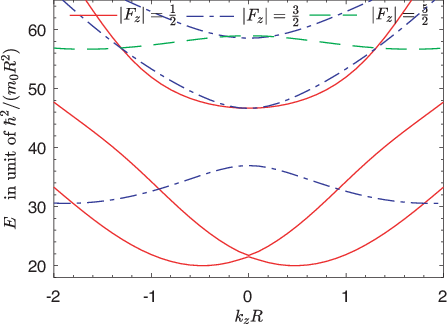}
\caption{\label{fig_subbands110}Hole subband dispersions for growth direction [110]. For a moderate nanowire radius $R=10$ nm, the energy unit is $\hbar^{2}/(m_{0}R^{2})\approx0.763$ meV. The energy gap at the anticrossing $k_{z}=0$ of the lowest two subband dispersions is about $0.23\hbar^{2}/(m_{0}R^{2})$. The band minima are located at $k_{z}R\approx\pm0.5$. Each dispersion line is two-fold (spin) degenerate.}
\end{figure}

The axial approximation gives rise to highly accurate subband dispersions for both growth directions [001] and [111] (see Ref.~\cite{Li-lirui:2024we}). Here we focus on another representative growth direction [110]. When the nanowire growth direction is along the [110] crystal axis, i.e., $k_{x}\parallel[00\bar{1}]$, $k_{y}\parallel[\bar{1}10]$, and $k_{z}\parallel[110]$, we have $\tilde{\gamma}_{1}=(\gamma_{2}+3\gamma_{3})/4$, $\tilde{\gamma}_{2}=(\gamma_{2}+\gamma_{3})/2$, and $\tilde{\gamma}_{3}=(3\gamma_{2}+5\gamma_{3})/8$ by setting $\theta=\pi/2$ and $\phi=\pi/4$ in Eq.~(\ref{eq_gamma})~\cite{PhysRevB.52.11132}. The hole subband dispersions for growth direction [110] are shown in Fig.~\ref{fig_subbands110}. The energy gap at the anticrossing of the lowest two subband dispersions is only about $0.23\hbar^{2}/(mR^{2})$, even smaller than that (about $0.61\hbar^{2}/(mR^{2})$) for growth direction [111]~\cite{Li-lirui:2024we}. The band minima are located at $k_{z}R\approx\pm0.50$.

The difference between $H_{\rm LK}$ and $H^{\rm ax}_{\rm LK}$ for growth direction [110] reads~\cite{winkler2003spin}
\begin{eqnarray}
H'_{110}&=&-\frac{3\hbar^{2}}{96m_{0}}(\gamma_{3}-\gamma_{2})\Big(6(k^{2}_{+}+k^{2}_{-})(J^{2}_{z}-5/4)\nonumber\\
&&+16k_{z}k_{+}\{J_{z},J_{+}\}+16k_{z}k_{-}\{J_{z},J_{-}\}\nonumber\\
&&-2(k_{+}k_{-}-2k^{2}_{z})(J^{2}_{+}+J^{2}_{-})-3(k^{2}_{+}J^{2}_{+}+k^{2}_{-}J^{2}_{-})\Big).\nonumber\\
\end{eqnarray}
Detailed calculations show $H'_{110}$ has perturbation matrix elements between the eigenstates of $|F_{z}|=1/2$ and $|F_{z}|\ge3/2$. Hence, the axial approximation is less accurate for growth direction [110] than growth directions [100] and [111]. Also, the crossings between the dispersion lines of $|F_{z}|=1/2$ and $|F_{z}|=3/2$ in Fig.~\ref{fig_subbands110} will become anticrossings when $H'_{110}$ is taken into consideration~\cite{RL2023c}.

\begin{figure}
\includegraphics{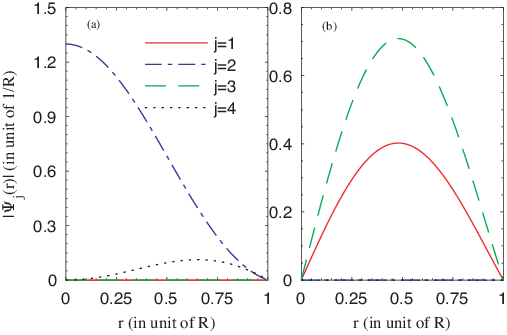}
\caption{\label{fig_eigenfuns110}The ground state (a) and the first excited state (b) wavefunctions at $k_{z}=0$ for growth direction [110]. The four components $\Psi_{1,2,3,4}(r)$ of the wavefunction~(\ref{Eq_eigenfunc}) are plotted as a function of the coordinate $r$.}
\end{figure}

The ground and the first excited state wavefunctions at $k_{z}=0$ are shown in Figs.~\ref{fig_eigenfuns110}(a) and (b), respectively. We find growth direction [110] has normal sequence of ground and first excited states similar to that of growth direction [001]~\cite{Li-lirui:2024we}. While growth direction [111] has inverted sequence of ground and first excited states. Roughly speaking, the ground state of growth directions [001] and [110] becomes the first excited state of growth direction [111]~\cite{Li-lirui:2024we}.

\begin{figure}
\includegraphics{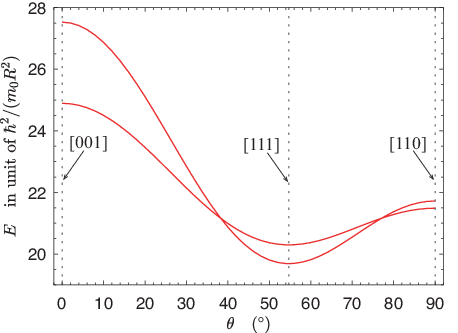}
\caption{\label{fig_levelcrossing}The lowest two subband energies at $k_{z}=0$ as a function of the growth direction $\theta$. Here $\phi$ is fixed at $\phi=45^{\circ}$. The energy gap closes both at $\theta\approx38.258^{\circ}$ and at $\theta\approx77.069^{\circ}$.}
\end{figure}

Quantum state inversion has been extensively studied in topological insulator~\cite{doi:10.1126/science.1133734,RevModPhys.83.1057}, and state inversion usually indicates the existence of gap closing site. We anticipate there are gap closing sites between growth directions [001]/[110] and [111]. When one angle of the growth direction is fixed at $\phi=45^{\circ}$, and the other angle $\theta$ is varied from $0^{\circ}$ to $90^{\circ}$, i.e., from growth direction [001] to [110], the lowest two subband energies at $k_{z}=0$ as a function of $\theta$ are shown in Fig.~\ref{fig_levelcrossing}. We indeed find two gap closing sites at $\theta\approx38.258^{\circ}$~\cite{Li-lirui:2024we} and $\theta\approx77.069^{\circ}$.

\section{\label{sec_splittingmagnetic}Spin splitting in a strong magnetic field}
{\color{red}}
The lowest two subband dispersions, e.g., the lowest two solid lines shown in Fig.~\ref{fig_subbands110}, can be regarded as two shifted parabolic curves with an anticrossing at $k_{z}=0$. Especially, the energy gap at the anticrossing is rather small for both growth directions [111] and [110] (See Fig.~\ref{fig_levelcrossing}). This peculiar low-energy subband structure indicates the existence of a strong `spin' (pseudo spin)-orbit coupling~\cite{RL2023a,RL2023b}, where the `spin'  is introduced to describe the two shifted parabolic curves. In order to make use of this strong `spin'-orbit coupling, we need to lift the spin degeneracy in the subband dispersions. This can be simply achieved by a strong magnetic field, such that two sets of combined dispersions with strong `spin'-orbit coupling and isolated from each other are obtainable~\cite{RL2023b}. 

The magnetic field can be applied either longitudinally or transversely. The magnetic field enters into the effective mass Hamiltonian via the vector potential~\cite{Semina:2015aa,PhysRevB.84.195314}, the effective mass Hamiltonian now reads 
\begin{equation}
H=H^{\rm ax}_{\rm LK}({\boldsymbol k}\rightarrow{\boldsymbol k'}={\boldsymbol k}+e{\bf A}/\hbar)+V(r)+2\kappa\mu_{B}{\bf B}\cdot{\bf J},\label{eq_efmassmagnetic}
\end{equation}
where ${\bf A}=(-B_{z}y/2,B_{z}x/2,0)$ when the field is longitudinal and ${\bf A}=(0,0,B_{x}y)$ when the field is transverse, $\kappa=3.41$ is the Luttinger magnetic parameter of semiconductor Ge~\cite{PhysRevB.4.3460}. We have properly chosen the vector potential such that $k_{z}$ in Hamiltonian (\ref{eq_efmassmagnetic}) is still conserved. In the following, we use perturbation theory to calculate the spin splitting in the subband dispersions, and Hamiltonian (\ref{eq_efmassmagnetic}) can be written as two parts
\begin{equation}
H=H_{0}+H_{B},\label{eq_efmassmagnetic2}
\end{equation}
where $H_{0}$ is zeroth order term given by Eq.~(\ref{eq_efmassHamil}) and the perturbation term $H_{B}$ contains both the Zeeman term and the orbital terms of the magnetic field.  The detailed form of $H_{B}$ is given in appendix~\ref{Appendix_c}. 

As we have shown in Sec.~\ref{sec_traneq}, both the eigenvalues and the corresponding eifenfunctions of $H_{0}$ are exactly obtainable. Now, we calculate the magnetic field induced spin splittings in the lowest two subband dispersions. Because the energy gap at the anticrossing is rather small for both growth directions [111] and [110], here we must use quasi-degenerate perturbation theory, i.e., Hamiltonian (\ref{eq_efmassmagnetic}) is written as a $4\times4$ matrix in the quasi degenerate Hilbert subspace spanned by the lowest four eigenstates. The detailed  procedure of calculations using the spherical approximation can be found in Ref.~\cite{RL2022a}. Diagonalizing the $4\times4$ matrix, we obtain the subband dispersions with spin splitting. 

\begin{figure}
\includegraphics{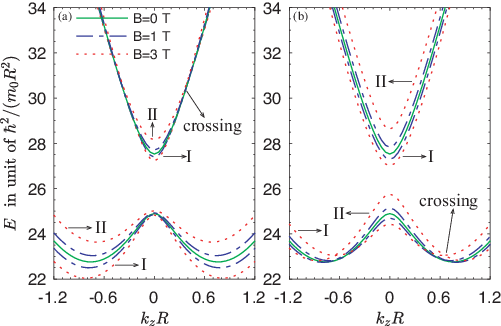}
\caption{\label{fig_spinsplit001}Spin splittings in the lowest two subband dispersions for growth direction [001]. The results of the strong longitudinal field (a) and the strong transverse field (b).}
\end{figure}

\begin{figure}
\includegraphics{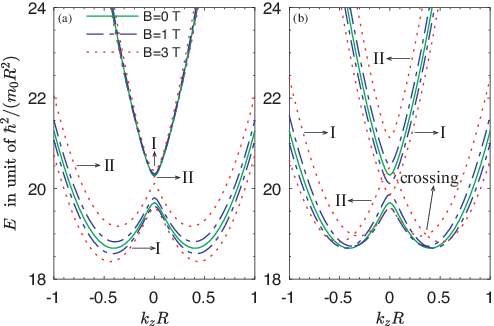}
\caption{\label{fig_spinsplit111}Spin splittings in the lowest two subband dispersions for growth direction [111]. The results of the strong longitudinal field (a) and the strong transverse field (b).}
\end{figure}

\begin{figure}
\includegraphics{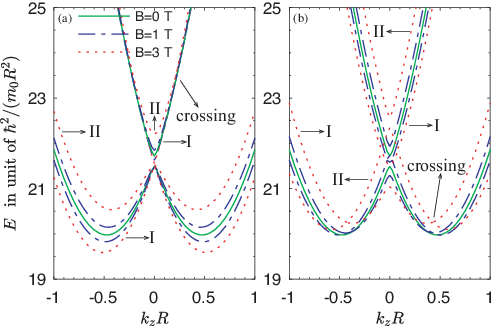}
\caption{\label{fig_spinsplit110}Spin splittings in the lowest two subband dispersions for growth direction [110]. The results of the strong longitudinal field (a) and the strong transverse field (b).}
\end{figure}

In Figs.~\ref{fig_spinsplit001}, \ref{fig_spinsplit111}, and \ref{fig_spinsplit110}, we show the spin splittings in the lowest two subband dispersions for growth directions [001], [111], and [110], respectively. Also, the results of the longitudinal field and the transverse field are shown in panel (a) and panel (b) of each figure, respectively. We can see subband crossings in Figs.~{\ref{fig_spinsplit001}}, \ref{fig_spinsplit111}, and \ref{fig_spinsplit110}, and these crossings imply the possibility of classifying the dispersion lines into different spin sets, e.g., one set I belongs to spin down and the other set II belongs to spin up. Each set I or II can be modelled by a two-band Hamiltonian~\cite{RL2023b}
\begin{equation}
 H^{\rm ef}=\hbar^{2}k^{2}_{z}/(2m^{*}_{h})+\alpha\sigma^{x}k_{z}+g^{*}_{h}\mu_{B}B\sigma^{z}/2,\label{eq_twobanddes}
 \end{equation}
where $\sigma^{x,z}$ are Pauli `spin' matrices. The `spin'-orbit coupling $\alpha$ and the effective $g$-factor $g^{*}_{h}$ are magnetic field dependent~\cite{RL2023a}. The two-band description of each set is further confirmed in Sec.~\ref{sec_effhamiltonian} via constructing the low-energy effective Hamiltonian of the hole gas. 

We note that, in terms of the real hole spin, a similar Hamiltonian like Eq.~(\ref{eq_twobanddes}) is also obtainable by using a strong electric field to lift the spin-degeneracy~\cite{PhysRevB.84.195314,PhysRevB.90.195421,PhysRevB.105.075308}. The difference between the `spin' (pseudo spin) and real spin is obvious, the real spin splitting is the splitting between lines I and II in Figs.~\ref{fig_spinsplit001}, \ref{fig_spinsplit111}, and \ref{fig_spinsplit110}. Also, the longitudinal $g$-factor of the real hole spin is rather small, about 0.14~\cite{PhysRevB.84.195314,PhysRevB.87.161305,PhysRevB.93.121408} calculated using the spherical approximation. While the longitudinal $g$-factor of the hole `spin' can be very large in the weak magnetic field region~\cite{RL2023a,RL2023b}.
 
Let us discuss the differences between the dispersions of the three growth directions. The energy gap at the anticrossing is positive for both growth directions [001] and [110], while it is negative for growth direction [111] (see also Fig.~\ref{fig_levelcrossing}). When the magnetic field is longitudinal, the spin splitting at $k_{z}=0$ of the lowest subband dispersion is much smaller than that of the second lowest subband dispersion for both [001] and [110] growth directions (see Figs.~\ref{fig_spinsplit001}(a) and \ref{fig_spinsplit110}(a)), while this result becomes reversed for  [111] growth direction (see Fig.~\ref{fig_spinsplit111}(a)). Also, the magnitude of the energy gap at the anticrossing is rather small for both growth directions [111] and [110].

\section{\label{sec_effhamiltonian}Low-energy effective Hamiltonian}
{\color{red}}
We now construct the low-energy effective Hamiltonians that qualitatively describe both the lowest two subband dispersions (see Fig.~\ref{fig_subbands110}) and their spin splittings in a strong magnetic field (see Figs.~{\ref{fig_spinsplit001}}, \ref{fig_spinsplit111}, and \ref{fig_spinsplit110}). We first focus on the absent magnetic field case, the low-energy effective Hamiltonian can be constructed as follows. The $k_{z}$ terms in Hamiltonian $H_{0}$ (\ref{eq_efmassHamil}) are treated as perturbation, e.g., we write $H_{0}=H_{0}(k_{z}=0)+H_{0}(k_{z}\neq0)$. The lowest four eigenstates of $H_{0}(k_{z}=0)$ are solved as $|e+\rangle$, $|e-\rangle$, $|g+\rangle$, and $|g-\rangle$ (see e.g., Fig. \ref{fig_eigenfuns110}). Then, $H_{0}$ can be written as an effective $4\times4$ Hamiltonian in the Hilbert subspace spanned by the above four eigenstates. In 2011, Kloeffel et al. obtained the following effective Hamiltonian based on the spherical approximation~\cite{PhysRevB.84.195314}
\begin{eqnarray}
H^{\rm ef}_{0}&=&\frac{\hbar^{2}k^{2}_{z}}{4}\left(\frac{1}{m^{*}_{e}}+\frac{1}{m^{*}_{g}}\right)+C\frac{\hbar^{2}}{m_{0}R}k_{z}\tau^{x}s^{x}\nonumber\\
&&+\left[\frac{\hbar^{2}k^{2}_{z}}{4}\left(\frac{1}{m^{*}_{e}}-\frac{1}{m^{*}_{g}}\right)+\frac{\Delta}{2}\frac{\hbar^{2}}{m_{0}R^{2}}\right]\tau^{z}.\label{eq_Hamiltoniankz}
\end{eqnarray}
where ${\boldsymbol \tau}$ and ${\bf s}$ are Pauli matrices defined in the orbital $\{g,e\}$ and spin $\{+,-\}$ subspaces, respectively, $\Delta$ is the energy spacing between $|e\pm\rangle$ and $|g\pm\rangle$. Note that the axial approximation still gives rise to Eq.~(\ref{eq_Hamiltoniankz}). The values of the parameters $m^{*}_{e,g}$, $C$, and $\Delta$ are given in Tab.~\ref{tab1}, and their analytical expressions are given in appendix~\ref{Appendix_d}.

\begin{table*}
  \centering 
  \caption{\label{tab1}The parameters of the effective Hamiltonians. These values are calculated using the formulas given in appendix~\ref{Appendix_d}, and those values given in the parentheses are obtained by a band fitting. The values based on the spherical and  the axial approximations are indicated by `sp' and  [001], [111], and [110], respectively.}
  
\begin{ruledtabular}
\begin{tabular}{cccccccccccccc}
&$m^{*}_{e}/m_{0}$\footnote{$m_{0}$ is the free electron mass}& $m^{*}_{g}/m_{0}$& $C$& $\Delta$& $Z_{1}$& $Z_{2}$& $Z_{3}$& $Z_{4}$& $X_{1}$& $X_{2}$& $X_{3}$& $X_{4}$&$U$\\
sp~\cite{PhysRevB.84.195314,RL2023b}&$0.054$ ($0.074$)& $0.043$~($0.074$)& $7.27$~($7.12$)& $0.77$& $0.75$& $0.82$& $-2.38$ ($-3.62$)& $0.65$& $2.73$& $-0.18$& $8.04$ ($4.66$)& $0.57$&$0.15$\\
$[001]$&$0.058$ ($0.092$)& $0.046$~($0.092$)& $8.83$~($8.50$)& $2.63$& $0.82$& $1.23$& $-2.76$ ($-4.15$)& $0.65$& $3.22$& $0.32$& $7.51$ ($5.12$)& $0.47$&$0.16$\\
$[111]$&$0.051$ ($0.066$)&$0.041$ ($0.066$)&$6.28$ ($6.22$)&$-0.61$&$0.75$&$0.46$&$-2.16$ ($-3.34$)&$0.65$&$2.35$&$-0.42$&$8.23$ ($6.52$) &$0.67$&$0.15$\\
$[110]$&$0.053$ ($0.071$)&$0.042$ ($0.071$)&$6.89$ ($6.77$)&$0.23$&$0.74$&$0.69$&$-2.30$ ($-3.29$)&$0.65$&$2.59$&$-0.27$&$8.13$ ($6.20$)&$0.61$&$0.15$
\end{tabular}
\end{ruledtabular}
\end{table*}

The effective Hamiltonian (\ref{eq_Hamiltoniankz}) indeed qualitatively reproduces the two shifted parabolic curves shown in Fig.~\ref{fig_subbands110}. However, both the site and the value of the subband minimum predicted by the effective Hamiltonian (\ref{eq_Hamiltoniankz}) are slightly different from the exact results given by Eq.~(\ref{eq_transc})~\cite{RL2023b}. This discrepancy is due to the fact that $H^{\rm ef}_{0}$ is valid only at small $k_{z}R$ ($k_{z}R\ll1$), while the exact site of the band minimum is located at $|k_{z}R|\approx0.4\rightarrow0.76$ depending on the growth direction. We also give the band fitting values of $m^{*}_{e,g}$ and $C$ in Tab.~\ref{tab1}~\cite{RL2023b}. The $k^{2}_{z}$ term in the second line of Eq.~(\ref{eq_Hamiltoniankz}) is much smaller than that in the first line, we thus neglect this term in the following. Hence, we have
\begin{equation}
H^{\rm ef}_{0}\approx\frac{\hbar^{2}k^{2}_{z}}{2m^{*}_{h}}+C\frac{\hbar^{2}}{m_{0}R}k_{z}\tau^{x}s^{x}+\frac{\Delta}{2}\frac{\hbar^{2}}{m_{0}R^{2}}\tau^{z},\label{eq_Hamiltoniankz2}
\end{equation}
where $1/m^{*}_{h}=(m^{*}_{e}+m^{*}_{g})/(2m^{*}_{e}m^{*}_{g})$. We also emphasize that the reduction from Eq.~(\ref{eq_Hamiltoniankz}) to Eq.~(\ref{eq_Hamiltoniankz2}) is an approximation, and does not mean $m^{*}_{e}=m^{*}_{g}$. 

Since the low-energy effective Hamiltonian governed by Hamiltonian $H_{0}$ is constructed successfully, we now move to construct the effective Hamiltonian of the magnetic term $H_{B}$ that gives rise to the spin splittings in Figs.~{\ref{fig_spinsplit001}}, \ref{fig_spinsplit111}, and \ref{fig_spinsplit110}. In the following, we separately consider the cases of the longitudinal field and the transverse field. In addition, the spin splitting induced by an  external electric field is also considered briefly. 

Note that if we want to keep the small $k^{2}_{z}$ term in the second line of Eq.~(\ref{eq_Hamiltoniankz}), the  classification of the subband dispersions into different spin sets is still feasible. We just need to add this term back to the Zeeman term in Eqs.~(\ref{eq_H_L1}), (\ref{eq_H_L2}), (\ref{eq_H_T1}), and (\ref{eq_H_T2}) given in the following.

\subsection{The strong longitudinal field case}
{\color{red}}
When the magnetic field is applied longitudinally, i.e., the field ${\bf B}=(0,0,B)$ is parallel to the nanowire axis, in the  low-energy Hilbert subspace mentioned above, the magnetic term $H_{B}$ in Eq.~(\ref{eq_efmassmagnetic2}) can also be written as an effective $4\times4$ Hamiltonian~\cite{PhysRevB.84.195314,RL2023b} 
\begin{equation}
H^{\rm ef}_{B}=\mu_{B}B(Z_{1}s^{z}+Z_{2}\tau^{z}s^{z}-Z_{3}k_{z}R\tau^{y}s^{y})+\frac{e^{2}B^{2}R^{2}}{2m_{0}}Z_{4}\tau^{z},
\end{equation}
where $Z_{i}$ ($i=1,2,3,4$) are parameters describing both the Zeeman and the orbital effects of the longitudinal field (see appendix~\ref{Appendix_d}).  Note that  the $Z_{4}$ term is caused by orbital effects of the magnetic field proportional to $B^{2}$, and cannot be neglected because here the magnetic field is strong. The values of $Z_{1,2,3,4}$ based on the spherical and the axial approximations are given in Tab.~\ref{tab1}. Interestingly, the total effective Hamiltonian $H^{\rm ef}=H^{\rm ef}_{0}+H^{\rm ef}_{B}$ is block diagonalized. In the Hilbert subspace spanned by $|e+\rangle$ and $|g-\rangle$, we have one effective two-band Hamiltonian~\cite{RL2023b} 
\begin{eqnarray}
H^{\rm ef}&=&\frac{\hbar^{2}k^{2}_{z}}{2m^{*}_{h}}+\left(C\frac{\hbar^{2}}{m_{0}R}+Z_{3}\mu_{B}BR\right)k_{z}\sigma^{x}\nonumber\\
&&+\left(\frac{\Delta}{2}\frac{\hbar^{2}}{m_{0}R^{2}}+Z_{1}\mu_{B}B+Z_{4}\frac{e^{2}B^{2}R^{2}}{2m_{0}}\right)\sigma^{z}.\label{eq_H_L1}
\end{eqnarray}
where $\sigma^{x}$ and $\sigma^{z}$ are Pauli matrices defined in this Hilbert subspace. In the Hilbert subspace spanned by $|e-\rangle$ and $|g+\rangle$, we have the other effective two-band Hamiltonian~\cite{RL2023b}  
\begin{eqnarray}
H^{\rm ef}&=&\frac{\hbar^{2}k^{2}_{z}}{2m^{*}_{h}}+\left(C\frac{\hbar^{2}}{m_{0}R}-Z_{3}\mu_{B}BR\right)k_{z}\sigma^{x}\nonumber\\
&&+\left(\frac{\Delta}{2}\frac{\hbar^{2}}{m_{0}R^{2}}-Z_{1}\mu_{B}B+Z_{4}\frac{e^{2}B^{2}R^{2}}{2m_{0}}\right)\sigma^{z}.\label{eq_H_L2}
\end{eqnarray}
These two equations (\ref{eq_H_L1}) and (\ref{eq_H_L2}) have the exact form as Eq.~(\ref{eq_twobanddes}), and nicely explain the two sets of the combined subband dispersions I and II shown in Figs.~\ref{fig_spinsplit001}(a), \ref{fig_spinsplit111}(a), and \ref{fig_spinsplit110}(a). The subband dispersions in a strong longitudinal field can indeed be classified into different spin sets!  Actually, the conservation of $F_{z}$ in model (\ref{eq_efmassmagnetic}) for the longitudinal field case also implies this result. Also, these two equations show that both `spin'-orbit coupling $\alpha$ and effective $g$-factor $g^{*}_{h}$ depend on the magnetic field $B$. 

\begin{figure}
\includegraphics{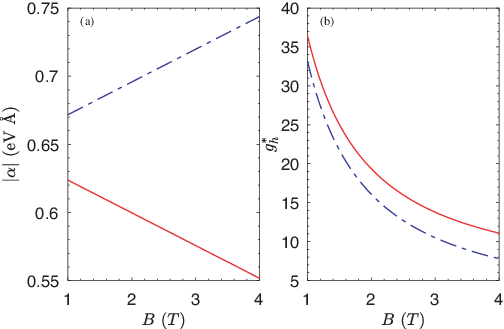}
\caption{\label{fig_socvsB001L}The `spin'-orbit coupling $\alpha$ (a) and the effective $g$-factor $g^{*}_{h}$ (b) as a function of the longitudinal field for growth direction [001]. The solid lines and the dashed-dot lines are results from Eqs.~(\ref{eq_H_L1}) and (\ref{eq_H_L2}), respectively. We have used the fitting parameters given in the parentheses of Tab.~\ref{tab1}, and $R=10$ nm.}
\end{figure}
\begin{figure}
\includegraphics{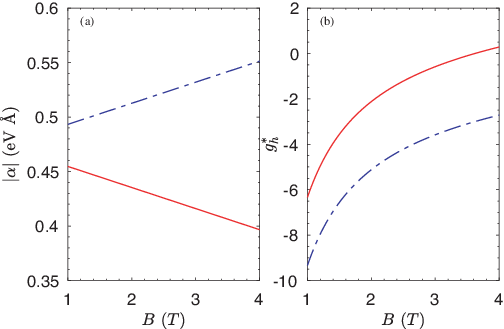}
\caption{\label{fig_socvsB111L}The `spin'-orbit coupling $\alpha$ (a) and the effective $g$-factor $g^{*}_{h}$ (b) as a function of the longitudinal field for growth direction [111]. The solid lines and the dashed-dot lines are results from Eqs.~(\ref{eq_H_L1}) and (\ref{eq_H_L2}), respectively. We have used the fitting parameters given in the parentheses of Tab.~\ref{tab1}, and $R=10$ nm.}
\end{figure}
\begin{figure}
\includegraphics{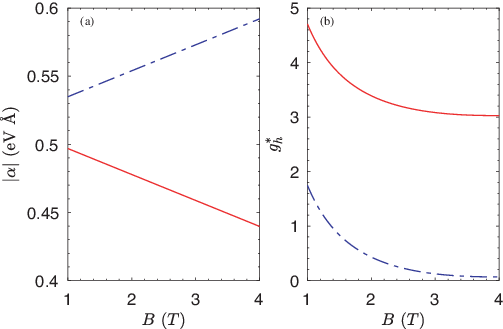}
\caption{\label{fig_socvsB110L}The `spin'-orbit coupling $\alpha$ (a) and the effective $g$-factor $g^{*}_{h}$ (b) as a function of the longitudinal field for growth direction [110]. The solid lines and the dashed-dot lines are results from Eqs.~(\ref{eq_H_L1}) and (\ref{eq_H_L2}), respectively. We have used the fitting parameters given in the parentheses of Tab.~\ref{tab1}, and $R=10$ nm.}
\end{figure}

Both the `spin'-orbit coupling $\alpha$ and the effective $g$-factor $g^{*}_{h}$ are plotted as a function of the longitudinal field $B$ in Figs.~\ref{fig_socvsB001L}, \ref{fig_socvsB111L}, and \ref{fig_socvsB110L} for growth directions [001], [111], and [110], respectively. In these figures, the dashed-dot lines are results of the set I combined dispersions, i.e., modeled by Eq.~(\ref{eq_H_L2}), and the solid lines are results of the set II combined dispersions, i.e., modeled by Eq.~(\ref{eq_H_L1}). Because the energy gap $\Delta$ is relatively large for growth direction [001], the effective $g$-factor $g^{*}_{h}$ is relatively large too (see Fig.~\ref{fig_socvsB001L}(b)). The energy gap $\Delta$ is negative for growth direction [111], such that the effective $g$-factor $g^{*}_{h}$ is mostly negative. Moreover, the effective $g$-factor $g^{*}_{h}$ of the set II combined dispersions (see Eq.~(\ref{eq_H_L1})) can be tuned from negative to zero, and even to positive with increasing the magnetic field (see the solid line in Fig.~\ref{fig_socvsB111L}(b)). The effective $g$-factor $g^{*}_{h}$ is smallest (the absolute value) for growth direction [110] in comparison with the other two growth directions (see Fig.~\ref{fig_socvsB110L}(b)).

The lowest combined hole subband dispersions, i.e., the set I lines shown in Figs.~\ref{fig_spinsplit001}(a), \ref{fig_spinsplit111}(a), and \ref{fig_spinsplit110}(a), are involved in most applications. We find nanowires grown along the [111] direction are especially useful for the spin qubit applications, because both the `spin'-orbit coupling $\alpha$ and the Zeeman splitting $|g^{*}_{h}|\mu_{B}B$ increase with the longitudinal field $B$ (see the dashed-dot lines shown in Fig.~\ref{fig_socvsB111L}).

\subsection{The strong transverse field case}
 {\color{red}}
When the magnetic field is applied transversely, i.e., the field ${\bf B}=(B,0,0)$ is perpendicular to the nanowire, the low-energy effective Hamiltonian of the magnetic term $H_{B}$ in Eq.~(\ref{eq_efmassmagnetic2}) now is written as~\cite{PhysRevB.84.195314,RL2023b} 
\begin{equation}
H^{\rm ef}_{B}=\mu_{B}B(X_{1}s^{x}+X_{2}\tau^{z}s^{x}+X_{3}k_{z}R\tau^{x})+X_{4}\frac{e^{2}B^{2}R^{2}}{2m_{0}}\tau^{z},
\end{equation}
where $X_{i}$ ($i=1,2,3,4$) are parameters describing both the Zeeman and the orbital effects of the transverse magnetic field (see appendix~\ref{Appendix_d}). The $X_{4}$ term is caused by terms proportional to $B^{2}$. The values of $X_{1,2,3,4}$ are given in Tab.~\ref{tab1}. Now, it is more easy to find the block diagonalized feature of the total effective Hamiltonian $H^{\rm ef}=H^{\rm ef}_{0}+H^{\rm ef}_{B}$ than that in the longitudinal field case, because the operator $s^{x}$ is a conserved quantity here. For $s^{x}=1$, the total effective Hamiltonian $H^{\rm ef}$ reads~\cite{RL2023b} 
\begin{eqnarray}
H^{\rm ef}&=&\frac{\hbar^{2}k^{2}_{z}}{2m^{*}_{h}}+\left(C\frac{\hbar^{2}}{m_{0}R}+X_{3}\mu_{B}BR\right)k_{z}\tau^{x}\nonumber\\
&&+\left(\frac{\Delta}{2}\frac{\hbar^{2}}{m_{0}R^{2}}+X_{2}\mu_{B}B+X_{4}\frac{e^{2}B^{2}R^{2}}{2m_{0}}\right)\tau^{z}.\label{eq_H_T1}
\end{eqnarray}
For $s^{x}=-1$, the total effective Hamiltonian $H^{\rm ef}$ reads~\cite{RL2023b} 
\begin{eqnarray}
H^{\rm ef}&=&\frac{\hbar^{2}k^{2}_{z}}{2m^{*}_{h}}-\left(C\frac{\hbar^{2}}{m_{0}R}-X_{3}\mu_{B}BR\right)k_{z}\tau^{x}\nonumber\\
&&+\left(\frac{\Delta}{2}\frac{\hbar^{2}}{m_{0}R^{2}}-X_{2}\mu_{B}B+X_{4}\frac{e^{2}B^{2}R^{2}}{2m_{0}}\right)\tau^{z}.\label{eq_H_T2}
\end{eqnarray}
When the magnetic field is transverse, we have still recovered Eq.~(\ref{eq_twobanddes}) successfully. The classification of subband dispersions into different spin sets is still feasible. These two equations (\ref{eq_H_T1}) and (\ref{eq_H_T2}) now nicely explain the two sets of the combined subband dispersions I and II shown in Figs.~\ref{fig_spinsplit001}(b), \ref{fig_spinsplit111}(b), and \ref{fig_spinsplit110}(b). 

\begin{figure}
\includegraphics{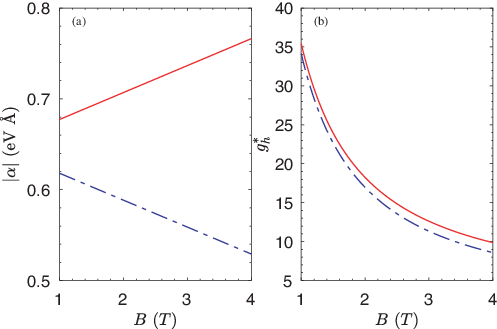}
\caption{\label{fig_socvsB001T}The `spin'-orbit coupling $\alpha$ (a) and the effective $g$-factor $g^{*}_{h}$ (b) as a function of the transverse field for growth direction [001]. The solid lines and the dashed-dot lines are results from Eqs.~(\ref{eq_H_T1}) and (\ref{eq_H_T2}), respectively. We have used the fitting parameters given in the parentheses of Tab.~\ref{tab1}, and $R=10$ nm.}
\end{figure}
\begin{figure}
\includegraphics{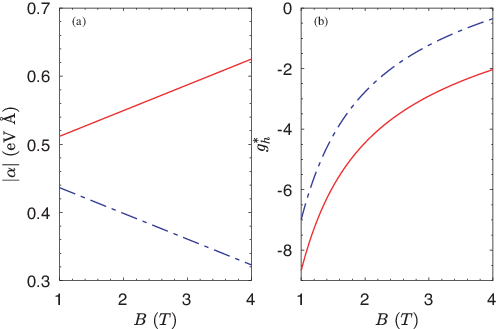}
\caption{\label{fig_socvsB111T}The `spin'-orbit coupling $\alpha$ (a) and the effective $g$-factor $g^{*}_{h}$ (b) as a function of the transverse field for growth direction [111]. The solid lines and the dashed-dot lines are results from Eqs.~(\ref{eq_H_T1}) and (\ref{eq_H_T2}), respectively. We have used the fitting parameters given in the parentheses of Tab.~\ref{tab1}, and $R=10$ nm.}
\end{figure}
\begin{figure}
\includegraphics{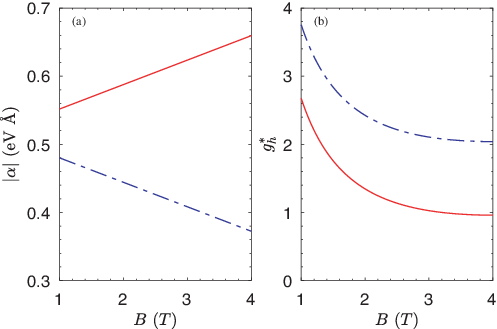}
\caption{\label{fig_socvsB110T}The `spin'-orbit coupling $\alpha$ (a) and the effective $g$-factor $g^{*}_{h}$ (b) as a function of the transverse field for growth direction [110]. The solid lines and the dashed-dot lines are results from Eqs.~(\ref{eq_H_T1}) and (\ref{eq_H_T2}), respectively. We have used the fitting parameters given in the parentheses of Tab.~\ref{tab1}, and $R=10$ nm.}
\end{figure}

Both the `spin'-orbit coupling $\alpha$ and the effective $g$-factor $g^{*}_{h}$ are plotted as a function of the transverse field $B$ in Figs.~\ref{fig_socvsB001T}, \ref{fig_socvsB111T}, and \ref{fig_socvsB110T} for growth directions [001], [111], and [110], respectively. In these figures, the dashed-dot lines are the results of the set I confined dispersions, i.e., modeled by Eq.~(\ref{eq_H_T2}), and the solid lines are the results of the set II combined dispersions, i.e., modeled by Eq.~(\ref{eq_H_T1}). The results in the transverse field case are very similar to that in the longitudinal field case, i.e., both $\alpha$ and $g^{*}_{h}$ here have the same order of magnitudes as that in the longitudinal field case. The key difference is that, the two sets of combined subband dispersions in the transverse field case are not well separated from each other [see Figs.~\ref{fig_spinsplit001}(b), \ref{fig_spinsplit111}(b), \ref{fig_spinsplit110}(b)], while they are well separated in the longitudinal field case. Therefore, the potential applications of the two-band Hamiltonians~(\ref{eq_H_T1}) and (\ref{eq_H_T2}) in a transverse field may be limited by this unsuccessful separation.

\subsection{The transverse electric field case}
{\color{red}}
A transverse external electric field can certainly lift the spin degeneracy in the hole subband dispersions. The transverse electric field breaks the space-inversion symmetry in the nanowire cross-section, such that the so-called Rashba spin splitting occurs at the wave vector sites $k_{z}\neq0$~\cite{bychkov1984oscillatory,winkler2003spin,PhysRevB.62.4245}. Many papers have studied theoretically this effect in nanowires with various cross-sections~\cite{PhysRevB.84.195314,PhysRevB.97.235422,Watzinger:2016aa,PhysRevB.103.195444,PhysRevApplied.18.044038,PhysRevB.104.115425}. When an external electric field $(E_{x},0,0)$ is applied in the $x$ direction, holes in the nanowire feel an additional electric potential energy with Hamiltonian
\begin{equation}
H_{\rm ed}=-eE_{x}x=-eE_{x}r\cos\varphi.
\end{equation}
The hole spin-orbit coupling is implicitly contained in the multi-band Hamiltonian $H=H_{0}+H_{\rm ed}$, such that we shouldn't add an additional spin-orbit coupling term such as $\alpha_{h}{\bf E}\cdot({\bf k}\times{\bf J})$ to the multi-band Hamiltonian. In the low-energy Hilbert subspace spanned by $|e+\rangle$, $|e-\rangle$, $|g+\rangle$, and $|g-\rangle$, we write $H_{\rm ed}$ as an effective Hamiltonian~\cite{PhysRevB.84.195314}
\begin{equation}
H^{\rm ef}_{\rm ed}=eE_{x}RU\tau^{y}s^{z}.
\end{equation}
The analytical formula of parameter $U$ is given in appendix~\ref{Appendix_d} and its values calculated using both the spherical and the axial Luttinger-Kohn Hamiltonians are given in Tab.~\ref{tab1}.  

\begin{figure}
\includegraphics{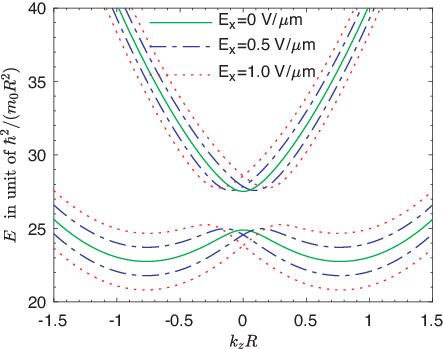}
\caption{\label{fig_spinsplitE001}Spin splittings induced by external electric field in the lowest two subband dispersions for growth direction [001]. We have used the fitting parameters given in the parentheses of Tab.~\ref{tab1}, and $R=10$ nm.}
\end{figure}
\begin{figure}
\includegraphics{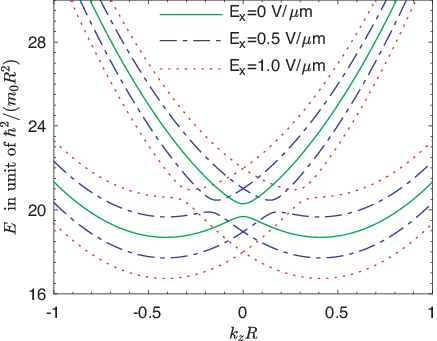}
\caption{\label{fig_spinsplitE111}Spin splittings induced by external electric field in the lowest two subband dispersions for growth direction [111]. We have used the fitting parameters given in the parentheses of Tab.~\ref{tab1}, and $R=10$ nm.}
\end{figure}
\begin{figure}
\includegraphics{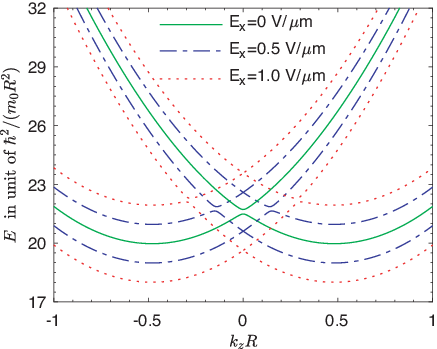}
\caption{\label{fig_spinsplitE110}Spin splittings induced by external electric field in the lowest two subband dispersions for growth direction [110]. We have used the fitting parameters given in the parentheses of Tab.~\ref{tab1}, and $R=10$ nm.}
\end{figure}

The  spin splittings induced by external electric field in the lowest two subband dispersions can be obtained by directly diagonalizing the $4\times4$ effective Hamiltonian $H^{\rm ef}=H^{\rm ef}_{0}+H^{\rm ef}_{\rm ed}$. The results of the spherical approximation were shown in Ref.~\cite{PhysRevB.84.195314}. The results of the axial approximation are shown in Figs.~\ref{fig_spinsplitE001}, \ref{fig_spinsplitE111}, and \ref{fig_spinsplitE110} for growth directions [001], [111], and [110], respectively. At a first glance, the hole spin splitting induced by external electric field in quasi-one-dimension looks very unusual. It is apparently different from its quasi-two dimensional counterpart, that is characterizable by terms such as $(k^{3}_{+}\sigma_{-}-k^{3}_{-}\sigma_{+})E_{z}$~\cite{winkler2003spin,PhysRevB.62.4245}. It is also different from the spin splitting of the conduction-band electrons, where is characterized by a linear-in-momentum term $(k_{-}\sigma_{+}-k_{+}\sigma_{-})E_{z}$~\cite{bychkov1984oscillatory}. The essential reason of these differences is that the hole subband minimums in quasi-one-dimension are not located at  $k_{z}=0$, the $k_{z}=0$ is an energy anticrossing site instead. However, the structure inversion asymmetry induced by external electric field must give rise to spin splitting at the wave vectors $k_{z}\neq0$, such that strange looks of the subband dispersions with spin splitting appear in Figs.~\ref{fig_spinsplitE001}, \ref{fig_spinsplitE111}, and \ref{fig_spinsplitE110}. One can find the time-reversal symmetry is indeed hold in these figures. 

For very large transverse electric fields, it is possible to describe the lowest two subband dispersions, e.g. the two shifted parabolic curves of $E_{x}=1.0$ V/$\mu$m in Fig.~\ref{fig_spinsplitE110}, by an effective two-band Hamiltonian. Also, the spin-splitting now is characterizable by a linear-in momentum term $k_{z}\sigma^{x}E_{x}$~\cite{PhysRevB.90.195421}. However, for small transverse electric fields, a four-band effective Hamiltonian $H^{\rm ef}=H^{\rm ef}_{0}+H^{\rm ef}_{\rm ed}$ is necessary.

\section{\label{sec_summary}Discussion and summary}
{\color{red}}
A series of experiments have shown signs of very strong hole spin-orbit coupling in the Ge/Si core/shell nanowire heterostructure~\cite{PhysRevLett.101.186802,PhysRevResearch.3.013081}, spin-orbit length as small as 20 nm~\cite{PhysRevLett.112.216806} and even smaller value of 3 nm~\cite{Froning:2021aa} are reported. Also, the longitudinal $g$-factor of hole is reported to be comparable with the transverse $g$-factor~\cite{Vries:2018aa,Froning:2021aa}. We remark that these observations are explainable in terms of the hole `spin' introduced in this paper. While in terms of the real hole spin, the longitudinal $g$-factor should be much smaller than the transverse $g$-factor~\cite{PhysRevB.84.195314,PhysRevB.93.121408}. 

The `spin'-orbit coupling stems from the peculiar double-well anticrossing subband structure of the quasi-one-dimensional hole gas. If a strain term proportional to $J^{2}_{z}$~\cite{PhysRevB.90.115419,PhysRevB.105.075308,PhysRevB.106.235408} is added to Hamiltonian (\ref{eq_efmassHamil}), the low-energy double-well subband structure will disappear. The lowest subband dispersion will become a single-well parabolic curve, and the `spin'-orbit coupling is destroyed as well. In this sense, the theoretical results of our paper are suitable to relaxed nanowires.

In summary, we have outlined both the theoretical method in calculating the subband dispersions and the low-energy effective description of the quasi-one-dimensional hole gas. A series of analytical formulas are derived in terms of the Luttinger-Kohn Hamiltonian in the axial approximation. The axial approximation is most accurate for both growth directions [001] and [111]. It is feasible to obtain two sets of combined dispersions with strong `spin'-orbit coupling and isolated from each other by lifting the spin degeneracy in the subband dispersions. We further construct the low-energy effective Hamiltonians of the hole gas both in the absence and presence of the external magnetic field. Both the double-well anticrossing subband structure and the classification of the subband dispersions into spin sets are qualitatively captured in these effective Hamiltonians.

Although the two sets of the combined dispersions are not well separated from each other in a strong transverse field, they are indeed well separated in a strong longitudinal field. Also, from the aspect of experimental implementation, it may be easier to apply a strong magnetic field along the nanowire than to produce a strong electric field across the nanowire. Anyway, we have provided an alternative way to achieve strong hole `spin'-orbit coupling. We expect this strong `spin'-orbit coupling has broad applications in quantum computing, e.g., the electrical manipulation of the hole spin and spin-cavity interaction etc. After all, the `spin' in our paper is essentially an orbital degree of freedom, such that the strong `spin'-orbit coupling may become senseless in those cases where the real spin is concerned.

\appendix
\begin{widetext}
\section{\label{Appendix_A}Matrices of operators $J_{x,y,z}$ and $\Gamma_{1,3}$}
In this paper, the matrices of operators $J_{x,y,z}$ and $\Gamma_{1,3}$ are as follows:
\begin{eqnarray}
J_{x}&=&\left(\begin{array}{cccc}
		0&\frac{\sqrt{3}}{2}&0&0\\
		\frac{\sqrt{3}}{2}&0&1&0\\
		0&1&0&\frac{\sqrt{3}}{2}\\
		0&0&\frac{\sqrt{3}}{2}&0
	\end{array}\right),~J_{y}=\left(\begin{array}{cccc}
	0&-i\frac{\sqrt{3}}{2}&0&0\\
	i\frac{\sqrt{3}}{2}&0&-i&0\\
	0&i&0&-i\frac{\sqrt{3}}{2}\\
	0&0&i\frac{\sqrt{3}}{2}&0
\end{array}\right),~J_{z}=\left(\begin{array}{cccc}
	\frac{3}{2}&0&0&0\\
	0&\frac{1}{2}&0&0\\
	0&0&-\frac{1}{2}&0\\
	0&0&0&-\frac{3}{2}
\end{array}
\right),\nonumber\\
\Gamma_{1}&=&\left(\begin{array}{cccc}0&0&i&0\\0&0&0&i\\-i&0&0&0\\0&-i&0&0\end{array}\right),~~~\Gamma_{3}=\left(\begin{array}{cccc}0&-i&0&0\\i&0&0&0\\0&0&0&i\\0&0&-i&0\end{array}\right).
\end{eqnarray}

\section{\label{Appendix_b}The eigenfunction and the equations of the coefficients}
The four components of the eigenfunction in the corresponding energy region $`\pm$' can be written as~\cite{Li-lirui:2024we} 
{\small}\begin{eqnarray}
\Psi_{1}(r)&=&\left(c_{1}\frac{2i\tilde{\gamma}_{2}k_{z}}{\tilde{\gamma}_{3}\mu_{1}}+c_{2}\frac{\tilde{\gamma}_{1}(\mu^{2}_{1}-2k^{2}_{z})\pm\chi_{\mu_{1}}}{\sqrt{3}\tilde{\gamma}_{3}\mu^{2}_{1}}\right)J_{m-1}(\mu_{1}\,r)+\left(c_{3}\frac{2i\tilde{\gamma}_{2}k_{z}}{\tilde{\gamma}_{3}\mu_{2}}+c_{4}\frac{\tilde{\gamma}_{1}(\mu^{2}_{2}-2k^{2}_{z})-\chi_{\mu_{2}}}{\sqrt{3}\tilde{\gamma}_{3}\mu^{2}_{2}}\right)J_{m-1}(\mu_{2}\,r),\nonumber\\
\Psi_{2}(r)&=&\left(c_{1}\frac{-\tilde{\gamma}_{1}(\mu^{2}_{1}-2k^{2}_{z})\pm\chi_{\mu_{1}}}{\sqrt{3}\tilde{\gamma}_{3}\mu^{2}_{1}}-c_{2}\frac{2i\tilde{\gamma}_{2}k_{z}}{\tilde{\gamma}_{3}\mu_{1}}\right)J_{m}(\mu_{1}\,r)+\left(c_{3}\frac{-\tilde{\gamma}_{1}(\mu^{2}_{2}-2k^{2}_{z})-\chi_{\mu_{2}}}{\sqrt{3}\tilde{\gamma}_{3}\mu^{2}_{2}}-c_{4}\frac{2i\tilde{\gamma}_{2}k_{z}}{\tilde{\gamma}_{3}\mu_{2}}\right)J_{m}(\mu_{2}\,r),\nonumber\\
\Psi_{3}(r)&=&c_{2}J_{m+1}(\mu_{1}\,r)+c_{4}J_{m+1}(\mu_{2}\,r),\nonumber\\
\Psi_{4}(r)&=&c_{1}J_{m+2}(\mu_{1}\,r)+c_{3}J_{m+2}(\mu_{2}\,r),\label{eq_wavefunction}
\end{eqnarray}
where $c_{1,2,3,4}$ are the expansion coefficients to be determined. The hard-wall boundary condition $\Psi(R,\varphi,z)=0$ can be written as~\cite{Li-lirui:2024we}
{\small
\begin{eqnarray}
\frac{2i\tilde{\gamma}_{2}k_{z}}{\tilde{\gamma}_{3}\mu_{1}}J_{m-1}(\mu_{1}R)c_{1}+\frac{\tilde{\gamma}_{1}(\mu^{2}_{1}-2k^{2}_{z})\pm\chi_{\mu_{1}}}{\sqrt{3}\tilde{\gamma}_{3}\mu^{2}_{1}}J_{m-1}(\mu_{1}R)c_{2}+\frac{2i\tilde{\gamma}_{2}k_{z}}{\tilde{\gamma}_{3}\mu_{2}}J_{m-1}(\mu_{2}R)c_{3}+\frac{\tilde{\gamma}_{1}(\mu^{2}_{2}-2k^{2}_{z})-\chi_{\mu_{2}}}{\sqrt{3}\tilde{\gamma}_{3}\mu^{2}_{2}}J_{m-1}(\mu_{2}R)c_{4}&=&0,\nonumber\\
\frac{-\tilde{\gamma}_{1}(\mu^{2}_{1}-2k^{2}_{z})\pm\chi_{\mu_{1}}}{\sqrt{3}\tilde{\gamma}_{3}\mu^{2}_{1}}J_{m}(\mu_{1}R)c_{1}-\frac{2i\tilde{\gamma}_{2}k_{z}}{\tilde{\gamma}_{3}\mu_{1}}J_{m}(\mu_{1}R)c_{2}-\frac{\tilde{\gamma}_{1}(\mu^{2}_{2}-2k^{2}_{z})+\chi_{\mu_{2}}}{\sqrt{3}\tilde{\gamma}_{3}\mu^{2}_{2}}J_{m}(\mu_{2}R)c_{3}-\frac{2i\tilde{\gamma}_{2}k_{z}}{\tilde{\gamma}_{3}\mu_{2}}J_{m}(\mu_{2}R)c_{4}&=&0,\nonumber\\
J_{m+1}(\mu_{1}R)c_{2}+J_{m+1}(\mu_{2}R)c_{4}&=&0,\nonumber\\
J_{m+2}(\mu_{1}R)c_{1}+J_{m+2}(\mu_{2}R)c_{3}&=&0.\nonumber\\\label{eq_boundarycondition}
\end{eqnarray}
}
The determinant of the coefficient matrix must be zero, such that we have the transcendental equations given by Eq.~(\ref{eq_transc}). At the site $k_{z}=0$, the boundary condition (\ref{eq_boundarycondition}) is reduced to
\begin{eqnarray}
\frac{\tilde{\gamma}_{1}+\sqrt{\tilde{\gamma}^{2}_{1}+3\tilde{\gamma}^{2}_{3}}}{\sqrt{3}\tilde{\gamma}_{3}}J_{m-1}(\mu_{1}R)c_{2}+\frac{\tilde{\gamma}_{1}-\sqrt{\tilde{\gamma}^{2}_{1}+3\tilde{\gamma}^{2}_{3}}}{\sqrt{3}\tilde{\gamma}_{3}}J_{m-1}(\mu_{2}R)c_{4}&=&0,\nonumber\\
\frac{-\tilde{\gamma}_{1}+\sqrt{\tilde{\gamma}^{2}_{1}+3\tilde{\gamma}^{2}_{3}}}{\sqrt{3}\tilde{\gamma}_{3}}J_{m}(\mu_{1}R)c_{1}-\frac{\tilde{\gamma}_{1}+\sqrt{\tilde{\gamma}^{2}_{1}+3\tilde{\gamma}^{2}_{3}}}{\sqrt{3}\tilde{\gamma}_{3}}J_{m}(\mu_{2}R)c_{3}&=&0,\nonumber\\
J_{m+1}(\mu_{1}R)c_{2}+J_{m+1}(\mu_{2}R)c_{4}&=&0,\nonumber\\
J_{m+2}(\mu_{1}R)c_{1}+J_{m+2}(\mu_{2}R)c_{3}&=&0.\label{eq_boundarycondition2}
\end{eqnarray}
One can see the equations of $c_{1}$ and $c_{3}$ are decoupled from the equations of $c_{2}$ and $c_{4}$. The equations of $c_{1}$ and $c_{3}$ leads to $\Psi_{1}(r)=\Psi_{3}(r)=0$, while the equations of $c_{2}$ and $c_{4}$ leads to $\Psi_{2}(r)=\Psi_{4}(r)=0$. Hence, at the site $k_{z}=0$, the eigenfunctions always have two vanishing components.

\section{\label{Appendix_c}The perturbation term $H_{B}$}
When a strong magnetic field ${\bf B}=(B_{x},0,B_{z})$ is applied, we make the following replacements in Eq.~(\ref{eq_efmassmagnetic})
\begin{equation}
k'_{x}=-i\partial_{x}-\frac{eB_{z}}{2\hbar}r\sin\varphi,~~~k'_{y}=-i\partial_{y}+\frac{eB_{z}}{2\hbar}r\sin\varphi,~~~k'_{z}=-i\partial_{z}+\frac{eB_{x}}{\hbar}r\sin\varphi.
\end{equation}
For obtaining the orbital effects of the magnetic field, we first calculate the following terms
\begin{eqnarray}
k'^{2}_{x}+k'^{2}_{y}&=&k^{2}_{x}+k^{2}_{y}-i\frac{eB_{z}}{\hbar}\partial_{\varphi}+\frac{eB^{2}_{z}r^{2}}{4\hbar^{2}},\nonumber\\
k'^{2}_{z}&=&k^{2}_{z}+2\frac{eB_{x}}{\hbar}k_{z}r\sin\varphi+\frac{e^{2}B^{2}_{x}}{\hbar^{2}}r^{2}\sin^{2}\varphi,\nonumber\\
\{k'_{z},k'_{-}\}&=&\{k_{z},k_{-}\}-\frac{eB_{x}}{\hbar}\left(\frac{1}{2}+\sin\varphi\,e^{-i\varphi}(ir\partial_{r}+\partial_{\varphi})\right)-i\frac{eB_{z}}{2\hbar}k_{z}re^{-i\varphi}-i\frac{e^{2}B_{x}B_{z}}{2\hbar^{2}}r^{2}\sin\varphi\,e^{-i\varphi},\nonumber\\
k'^{2}_{-}&=&k^{2}_{-}-\frac{eB_{z}}{\hbar}\left(r\partial_{r}e^{-2i\varphi}-ie^{-2i\varphi}\partial_{\varphi}\right)-\frac{e^{2}B^{2}_{z}r^{2}}{4\hbar^{2}}e^{-2i\varphi}.
\end{eqnarray}
Interested readers can also refer to Ref.~\cite{PhysRevB.97.235422}. Now the perturbation term $H_{B}$ in Eq.~(\ref{eq_efmassmagnetic2}) can be written as
\begin{equation}
H_{B}=\frac{\hbar^{2}}{2m_{0}}\left(\begin{array}{cccc}F'&S'&I'&0\\S'^{\dagger}&G'&0&I'\\I'^{\dagger}&0&G'&-S'\\0&I'^{\dagger}&-S'^{\dagger}&F'\end{array}\right)+2\kappa\mu_{B}{\bf B}\cdot{\bf J},\label{eq_app_H_B}
\end{equation}
where
\begin{eqnarray}
F'&=&(\gamma_{1}+\tilde{\gamma}_{1})\Big(-i\frac{eB_{z}}{\hbar}\partial_{\varphi}+\frac{e^{2}B^{2}_{z}r^{2}}{4\hbar^{2}}\Big)+(\gamma_{1}-2\tilde{\gamma}_{1})\Big(2\frac{eB_{x}}{\hbar}k_{z}r\sin\varphi+\frac{e^{2}B^{2}_{x}r^{2}}{\hbar^{2}}\sin^{2}\varphi\Big),\nonumber\\
G'&=&(\gamma_{1}-\tilde{\gamma}_{1})\Big(-i\frac{eB_{z}}{\hbar}\partial_{\varphi}+\frac{e^{2}B^{2}_{z}r^{2}}{4\hbar^{2}}\Big)+(\gamma_{1}+2\tilde{\gamma}_{1})\Big(2\frac{eB_{x}}{\hbar}k_{z}r\sin\varphi+\frac{e^{2}B^{2}_{x}r^{2}}{\hbar^{2}}\sin^{2}\varphi\Big),\nonumber\\
S'&=&-2\sqrt{3}\tilde{\gamma}_{2}\Big(-\frac{eB_{x}}{\hbar}\Big(\frac{1}{2}+\sin\varphi\,e^{-i\varphi}(ir\partial_{r}+\partial_{\varphi})\Big)-i\frac{eB_{z}}{2\hbar}k_{z}re^{-i\varphi}-i\frac{e^{2}B_{x}B_{z}}{2\hbar^{2}}r^{2}\sin\varphi\,e^{-i\varphi}\Big),\nonumber\\
I'&=&-\sqrt{3}\tilde{\gamma}_{3}\Big(-\frac{eB_{z}}{\hbar}\left(r\partial_{r}e^{-2i\varphi}-ie^{-2i\varphi}\partial_{\varphi}\right)-\frac{e^{2}B^{2}_{z}r^{2}}{4\hbar^{2}}e^{-2i\varphi}\Big).
\end{eqnarray}
For a longitudinal field we set $B_{x}=0$ and for a transverse field we set $B_{z}=0$ in Eq.~(\ref{eq_app_H_B}).

\section{\label{Appendix_d}The parameters of the low-energy effective Hamiltonian}
The low-energy Hilbert subspace is spanned by the lowest four eigenstates of $H_{0}(k_{z}=0)$, i.e.,
\begin{eqnarray}
|g-\rangle&=&\left(\begin{array}{c}0\\\Psi_{I,2}(r)\\0\\\Psi_{I,4}(r)e^{2i\varphi}\end{array}\right),~|g+\rangle=\left(\begin{array}{c}\Psi^{*}_{I,4}(r)e^{-2i\varphi}\\0\\\Psi^{*}_{I,2}(r)\\0\end{array}\right),~|e-\rangle=\left(\begin{array}{c}0\\\Psi^{*}_{II,3}(r)e^{-i\varphi}\\0\\\Psi^{*}_{II,1}(r)e^{i\varphi}\end{array}\right),~|e+\rangle=\left(\begin{array}{c}\Psi_{II,1}(r)e^{-i\varphi}\\0\\\Psi_{II,3}(r)e^{i\varphi}\\0\end{array}\right).
\end{eqnarray}
The corresponding coefficients $c_{1,2,3,4}$ in $\Psi_{I}(r)$ and $\Psi_{II}(r)$ are obtainable from Eq.~(\ref{eq_boundarycondition2}). Now, we can derive each term of the effective Hamiltonian, e.g., by projecting terms such as $H_{0}$, $H_{B}$, and $H_{\rm ed}$ onto the above low-energy Hilbert subspace. After some tedious algebra, the parameters of the low-energy effective Hamiltonian can be derived as
\begin{eqnarray}
m^{*}_{e}&=&\frac{m_{0}}{2\pi\int^{R}_{0}drr\Big((\gamma_{1}-2\tilde{\gamma}_{1})|\Psi_{II,1}(r)|^{2}+(\gamma_{1}+2\tilde{\gamma}_{1})|\Psi_{II,3}(r)|^{2}\Big)},\nonumber\\
m^{*}_{g}&=&\frac{m_{0}}{2\pi\int^{R}_{0}drr\Big((\gamma_{1}+2\tilde{\gamma}_{1})|\Psi_{I,2}(r)|^{2}+(\gamma_{1}-2\tilde{\gamma}_{1})|\Psi_{I,4}(r)|^{2}\Big)},\nonumber\\
C&=&\sqrt{3}\tilde{\gamma}_{2}R\int^{R}_{0}drr\int^{2\pi}_{0}d\varphi\Big(-\Psi^{*}_{I,2}(r)k_{+}\Psi_{II,1}(r)e^{-i\varphi}+\Psi^{*}_{I,4}(r)e^{-2i\varphi}k_{+}\Psi_{II,3}(r)e^{i\varphi}\Big),\nonumber\\
Z_{1}&=&2\pi\kappa\int^{R}_{0}drr\Big(\frac{3}{2}|\Psi_{II,1}(r)|^{2}-\frac{1}{2}|\Psi_{II,3}(r)|^{2}-\frac{1}{2}|\Psi_{I,2}(r)|^{2}+\frac{3}{2}|\Psi_{I,4}(r)|^{2}\Big)\nonumber\\
&&+\pi\int^{R}_{0}drr\Big(-(\gamma_{1}+\tilde{\gamma}_{1})|\Psi_{II,1}(r)|^{2}+(\gamma_{1}-\tilde{\gamma}_{1})|\Psi_{II,3}(r)|^{2}-2(\gamma_{1}+\tilde{\gamma}_{1})|\Psi_{I,4}(r)|^{2}\Big)\nonumber\\
&&+\pi\sqrt{3}\tilde{\gamma}_{3}\int^{R}_{0}drr\Big(\Psi^{*}_{II,1}(r)(r\partial_{r}+1)\Psi_{II,3}(r)-\Psi^{*}_{I,2}(r)(r\partial_{r}+2)\Psi_{I,4}(r)+c.c.\Big),\nonumber\\
Z_{2}&=&2\pi\kappa\int^{R}_{0}drr\Big(\frac{3}{2}|\Psi_{II,1}(r)|^{2}-\frac{1}{2}|\Psi_{II,3}(r)|^{2}+\frac{1}{2}|\Psi_{I,2}(r)|^{2}-\frac{3}{2}|\Psi_{I,4}(r)|^{2}\Big)\nonumber\\
&&+\pi\int^{R}_{0}drr\Big(-(\gamma_{1}+\tilde{\gamma}_{1})|\Psi_{II,1}(r)|^{2}+(\gamma_{1}-\tilde{\gamma}_{1})|\Psi_{II,3}(r)|^{2}+2(\gamma_{1}+\tilde{\gamma}_{1})|\Psi_{I,4}(r)|^{2}\Big)\nonumber\\
&&+\pi\sqrt{3}\tilde{\gamma}_{3}\int^{R}_{0}drr\Big(\Psi^{*}_{II,1}(r)(r\partial_{r}+1)\Psi_{II,3}(r)+\Psi^{*}_{I,2}(r)(r\partial_{r}+2)\Psi_{I,4}(r)+c.c.\Big),\nonumber\\
Z_{3}&=&(2\pi\sqrt{3}\tilde{\gamma}_{2}/R)\int^{R}_{0}dr\,r\Big(-ir\Psi^{*}_{I,2}(r)\Psi_{II,1}(r)+ir\Psi^{*}_{I,4}(r)\Psi_{II,3}(r)\Big),\nonumber\\
Z_{4}&=&\pi(\gamma_{1}+\frac{5}{2}\tilde{\gamma}_{1})\frac{1}{4R^{2}}\int^{R}_{0}drr^{3}\Big(|\Psi_{II,1}(r)|^{2}+|\Psi_{II,3}(r)|^{2}-|\Psi_{I,2}(r)|^{2}-|\Psi_{I,4}(r)|^{2}\Big)\nonumber\\
&&-\frac{\pi}{2R^{2}}\int^{R}_{0}drr^{3}\Big(-\frac{\sqrt{3}}{2}\tilde{\gamma}_{3}\big(\Psi^{*}_{II,1}(r)\Psi_{II,3}(r)-\Psi^{*}_{I,2}(r)\Psi_{I,4}(r)\big)-\frac{\sqrt{3}}{2}\tilde{\gamma}_{3}\big(\Psi^{*}_{II,3}(r)\Psi_{II,1}(r)-\Psi^{*}_{I,4}(r)\Psi_{I,2}(r)\big)\nonumber\\
&&+\frac{3}{4}\tilde{\gamma}_{1}\big(|\Psi_{II,1}(r)|^{2}-|\Psi_{I,4}(r)|^{2}\big)+\frac{7}{4}\tilde{\gamma}_{1}\big(|\Psi_{II,3}(r)|^{2}-|\Psi_{I,2}(r)|^{2}\big)\Big),\nonumber\\
X_{1}&=&2\pi\kappa\int^{R}_{0}drr\Big(\sqrt{3}\Psi^{*}_{II,1}(r)\Psi^{*}_{II,3}(r)+\Psi^{*}_{I,2}(r)\Psi^{*}_{I,2}(r)\Big)\nonumber\\
&&+2\pi\sqrt{3}\tilde{\gamma}_{2}\int^{R}_{0}drr\Big(\Psi^{*}_{II,1}(r)(r\partial_{r}+1)\Psi^{*}_{II,3}(r)-\Psi^{*}_{II,1}(r)\Psi^{*}_{II,3}(r)+\Psi^{*}_{I,2}(r)(r\partial_{r}+2)\Psi^{*}_{I,4}(r)\Big),\nonumber\\
X_{2}&=&2\pi\kappa\int^{R}_{0}drr\Big(\sqrt{3}\Psi^{*}_{II,1}(r)\Psi^{*}_{II,3}(r)-\Psi^{*}_{I,2}(r)\Psi^{*}_{I,2}(r)\Big)\nonumber\\
&&+2\pi\sqrt{3}\tilde{\gamma}_{2}\int^{R}_{0}drr\Big(\Psi^{*}_{II,1}(r)(r\partial_{r}+1)\Psi^{*}_{II,3}(r)-\Psi^{*}_{II,1}(r)\Psi^{*}_{II,3}(r)-\Psi^{*}_{I,2}(r)(r\partial_{r}+2)\Psi^{*}_{I,4}(r)\Big),\nonumber\\
X_{3}&=&-(2i\pi/R)\int^{R}_{0}drr^{2}\Big((\gamma_{1}+2\tilde{\gamma}_{1})\Psi^{*}_{I,2}(r)\Psi^{*}_{II,3}(r)+(\gamma_{1}-2\tilde{\gamma}_{1})\Psi^{*}_{I,4}(r)\Psi^{*}_{II,1}(r)\Big),\nonumber\\
X_{4}&=&\frac{\pi}{2R^{2}}\int^{R}_{0}drr^{3}\Big((\gamma_{1}-2\tilde{\gamma}_{1})\big(|\Psi_{II,1}(r)|^{2}-|\Psi_{I,4}(r)|^{2}\big)+(\gamma_{1}+2\tilde{\gamma}_{1})\big(|\Psi_{II,3}(r)|^{2}-|\Psi_{I,2}(r)|^{2}\big)\Big),\nonumber\\
U&=&-(i\pi/R)\int^{R}_{0}drr^{2}\Big(\Psi^{*}_{I,2}(r)\Psi^{*}_{II,3}(r)+\Psi^{*}_{I,4}(r)\Psi^{*}_{II,1}(r)\Big).
\end{eqnarray}
Here $c.c.$ means taking the complex conjugate of the preceding terms. 
\end{widetext}
\bibliography{Ref_Hole_spin}
\end{document}